\documentclass{fizik} 
\vol{}
\pyear{}
\received{}

\newcommand{\ben}{\begin{enumerate}}
\newcommand{\een}{\end{enumerate}}
\newcommand{\be}{\begin{equation}}
\newcommand{\ee}{\end{equation}}
\newcommand{\bse}{\begin{subequation}}
\newcommand{\ese}{\end{subequation}}
\newcommand{\bea}{\begin{eqnarray}}
\newcommand{\eea}{\end{eqnarray}}
\newcommand{\bc}{\begin{center}}
\newcommand{\ec}{\end{center}}

\newcommand{\n}{\noindent}

\def\frac#1#2{{\textstyle{#1\over #2}}}

\def\b#1{\kern-0.25pt\vbox{\hrule height 0.2pt\hbox{\vrule
width 0.2pt \kern2pt\vbox{\kern2pt \hbox{#1}\kern2pt}\kern2pt\vrule
width 0.2pt}\hrule height 0.2pt}}
\def\ST#1{\matrix{\vbox{#1}}}
\def\STrow#1{\hbox{#1}\kern-1.35pt}
\def\bv{\b{\phantom{1}}}

\def\eqalignD#1{
\vcenter{\openup1\jot\halign{
\hfil$\displaystyle{##}$~&
$\displaystyle{##}$\hfil~&
$\displaystyle{##}$\hfil\cr
#1}}
}

\def\text#1{\quad\hbox{#1}\quad}

\def\e{\epsilon}
\def\nuh{{\hat \nu}}
\def\muh{{\hat \mu}}

\def\E{\widehat {E}}

\def\W{\widehat {W}}
\def\rh{{\hat \rho}}
\def\lah{{\hat \lambda}}

\def\y{{\infty}}

\def\rw{\rightarrow}

\def\lra{\leftrightarrow}
\def\su{\widehat{su}}

\def\sp{\widehat{sp}}

\def\Nc{{\cal N}}

\author[BEGIN CUMMINS MATHIEU]
        {
        {\bf  Luc Bégin}\\
        {\it  Département de Physique, Université Laval}\\
        {\it  Québec, Canada G1K 7P4}\\
        {\bf  Chris Cummins}\\
        {\it  Mathematics Department, University of Concordia}\\
        {\it  Montréal Québec Canada H3G 1M8}\\
	{\bf  Pierre Mathieu}\\
        {\it  Département de Physique, Université Laval}\\
        {\it  Québec, Canada G1K 7P4}\\

        }
\title{Fusion bases for affine Lie algebras}

\begin{document}
\maketitle

\begin{abstract}
Fusion coefficients for affine Lie algebras
are fixed by the  corresponding tensor-product coefficients 
and a set of {\it threshold levels}. It is shown how the information 
concerning the 
threshold level is coded in the {\it fusion basis}, which is a 
set of inequalities that 
completely describes the fusion coefficients. The construction of the $\su(2)$
fusion basis is presented in detail.
\end{abstract}

\section{Fusion rules, tensor products and threshold levels}

We first introduce some basic concepts (and a bit of notation) in order to
prepare the ground for the formulation of fusion rules in terms of a fusion basis.

Fusion rules give the
number of independent couplings between three given primary fields in conformal
field theories. Here we are interested to those conformal
field theories having a Lie group
symmetry.  These are the Wess-Zumino-Witten models \cite{kz, GW}, whose
generating spectrum algebra is an affine Lie algebra at integer level. Their
primary fields  are in 1-1 correspondence with the integrable representations of
the appropriate affine Lie algebra at level
$k$. Denote this set by $P_+^{(k)}$ and a primary field by the corresponding
affine weight $\lah$.  Fusion coefficients ${\Nc_{\lah\muh}^{(k)}}~^{\nuh}$ are
defined by the product
\be  \lah\times \muh = \sum_{\nuh\in P_+^{(k)}} {\Nc_{\lah\muh}^{(k)}}~^{\nuh}
\; \nuh \ee To simplify the presentation, we consider only $\su(N)$.

An affine weight may be written as \be \lah=\sum_{i=0} ^{N-1} \lambda_{i}
{\widehat{\omega}}_{i}=[\lambda_0,\lambda_1,..., \lambda_{N-1}] \ee
where
${\widehat{\omega}}_{i}$ denote the fundamental weights of $\su(N)$. If
the Dynkin labels $\lambda_i$ are nonnegative, then the weight $\lah$  is the highest
weight of an integrable  representation of $\su(N)$ at level $k$, with  
 $k$  defined by $k=\sum_{i=0}^{N-1} \lambda_i$. 
To the affine
 weight $\lah$, we associate a finite weight ${\lambda}$ of the finite algebra
$su(N)$:
\be {\lambda}=\sum_{i=1}^{N-1} \lambda_i
{\omega}_{i} = (\lambda_1,...,\lambda_r) \ee where ${\omega}_{i}$  are the
fundamental weights of $su(N)$.  Thus $\lah$ is uniquely fixed from $\lambda$ and
$k$.

The fusion coefficient ${\Nc_{\lah\muh}^{(k)}}~^{\nuh}$ is fixed to a large
extent by the tensor-product coefficient of
the corresponding finite representations. 
We denote by ${\Nc_{\lambda\mu}}^{\nu}$ the multiplicity of the representation
$\nu$ in the tensor product $\lambda\otimes\mu$:
\be \lambda\otimes \mu = \sum_{\nu\in P_+} {\Nc_{\lambda\mu}}^{\nu}\;  \nu\ee
where by abuse of notation, we use the same symbol for the highest weight and
the  highest-weight representation. $P_+$ represents the set of integrable finite
weights. The precise relation between tensor-product and fusion-rule
coefficients is given by  the Kac-Walton formula ~\cite{wal, Kac, Furlan}:
\be {\Nc_{\lah\muh}^{(k)}}~^{\nuh}=
\sum_{w\in \W , ~ w\cdot{\hat\xi}=\nuh\in P_+^{(k)} }~{\Nc_{\lambda\mu}}^{\xi} {}~\epsilon(w) \label{kacwal}\ee $w$ is an element of
the affine Weyl group $\W$, of sign $\epsilon(w)$, and the dot indicates the shifted
action: $w\cdot\lah=w(\lah+\rh)-\rh$ where $\rh$ stands for the affine Weyl vector:
$
\rh=\sum_{i=0}^{N-1} {\widehat\omega}_{i}.$

The Kac-Walton formula can be transposed into a simple algorithm: one first
calculates the tensor product of the corresponding finite weights and then
extends every weight to its affine version at the appropriate value of $k$ and
shift-reflects back to the integrable affine sector those weights which have
negative zeroth Dynkin label. Weights that cannot be shift-reflected in the
integrable sector are ignored  
(for example this is  the case for those which have zeroth Dynkin label
equal to
$-1$). Here is a simple example: consider the $su(2)$ tensor-product \be (2)\otimes
(4)= (2)\oplus(4)\oplus(6)\ee and its affine extension at level 4:
\be [2,2]\times [0,4]= [2,2]+[0,4]+ [-2,6]\ee The last weight must be
reflected since it is not integrable: the shifted action of
$s_0$, the reflection with respect to the  zeroth affine root, is \be s_0\cdot
[-2,6]= s_0 ([-2,6]+[1,1])-[1,1]= [0,4]\ee
and this contributes with a minus sign ($\e(s_0)=-1$), cancelling then the other
$[0,4]$ representation; we thus find: $[2,2]\times [0,4]= [2,2]$.  On the other
hand, at level 5 the affine extension of the same product becomes
\be [3,2]\times [1,4]= [3,2]+[1,4]+[-1,6]\ee
The last weight is thus ignored and the final result is $[3,2]\times [1,4]=
[3,2]+[1,4]$.  For $k>5$ it is clear that there are no truncations, hence no
difference between the fusion coefficients and the tensor products. Moreover,
we see that the representation $(4)$ occurs at level 5 and higher.   
  We then say that its {\it threshold
level}, denoted by $k_0$, is 
$5$. 
The threshold level is thus the smallest
value of $k$ such that the fusion coefficient ${\Nc_{\lah\muh}^{(k)}}~^{\nuh}$ is
non-zero. If we indicate the threshold level by a subindex, we can write
\be (2)\otimes
(4)= (2)_4\oplus(4)_5\oplus(6)_6 \ee To read off a fusion at fixed level $k$, we
only keep terms with index not greater than $k$.  The concept of threshold level
was first introduced in  \cite{CMW}. It implies directly the inequality
\be {\Nc_{\lah\muh}^{(k)}}~^{\nuh} \leq
{\Nc_{\lah\muh}^{(k+1)}}~^{\nuh} \label{truc} \ee which in turn yields 
\be \lim_{k \rightarrow \y} {\Nc_{\lah\muh}^{(k)}}~^{\nuh}= 
{\Nc_{\lambda\mu}}^{\nu}.
\label{limformula} \ee

There are simple combinatorial methods that can be used for calculating
$su(N)$ tensor products, for instance, the Littlewood-Richardson (LR)  rule.  We
can then ask: can we read off the threshold level of a coupling from its LR
tableau?

\section{Tensor products, linear inequalities and elementary couplings}

Integrable weights in $su(N)$ can be represented by tableaux: the weight
$(\lambda_1,\lambda_2,
\cdots ,\lambda_{N-1})$ is associated to a left-justified tableau of $N-1$ rows 
with $\lambda_1+\lambda_2+\cdots +\lambda_{N-1}$ boxes in the first row,
$\lambda_1+\lambda_2+\cdots +\lambda_{N-2}$ boxes in the second row, etc. 
Equivalently, the tableau has $\lambda_1$ columns of 1 box, $\lambda_2$ columns
of 2 boxes, etc. The scalar representation has no boxes, or equivalently, any
number of columns of $N$ boxes. For instance, the
$su(3)$ weight $(1,1)$ is associated to the tableau
$\ST{\STrow{\bv\bv}\STrow{\bv}}$ and the $su(4)$ weight (2,3,0) to
$\ST{\STrow{\bv\bv\bv\bv\bv}\STrow{\bv\bv\bv}}$.

The  Littlewood-Richardson
rule is a simple combinatorial description of the tensor product of two $su(N)$
representations $\lambda\otimes \mu$.   The second tableau ($\mu$) is filled with
numbers as follows: the first row with
$1$'s, the second row with $2$'s, etc. All the boxes with a $1$ are then added  to
the first tableau according to 
following restrictions: (1) the resulting tableau must be regular: the number of 
boxes in a
given row must be smaller than or equal to the number of boxes in the row 
immediately above; (2) the resulting tableau must not contain two boxes 
marked by $1$
in the same column.  All the boxes marked by a $2$ are the added to
the resulting tableaux according to the above two rules (with $1$
is replaced by $2$) and the further restriction: (3) in counting from right to
left and top to bottom, the  number of
$1$'s must always be greater or equal to the number of $2$'s.
The process is repeated with the boxes marked by a $3, 4, \cdots, N-1$, with
the additional
rule that the number of
$i$'s must always be greater or equal to the number of $i+1$'s when counted from
right to left and top to bottom.
 The resulting Littlewood-Richardson (LR) tableaux are the Young
tableaux of the irreducible representations occurring in the decomposition.

Here is a simple $su(3)$ example: $(1,1)\otimes(1,1)\supset 2(1,1)$ since we can
draw two LR tableaux with  shape $(1,1)$ and an extra column of three boxes (the
total number of boxes being preserved, the resulting LR tableau must have 6
boxes):
\be \ST{\STrow{\bv\bv\b1}\STrow{\bv\b2}\STrow{\b1}}
\quad\quad\ST{\STrow{\bv\bv\b1}\STrow{\bv\b1}\STrow{\b2}} 
\ee

These rules can be rephrased in an algebraic way as follows \cite{CCS}. Define
$n_{ij}$ to be the number of boxes $i$ that appear in the LR tableau in the row
$j$. The LR conditions read:
\be \lambda_{j-1}+\sum_{i=1}^{k-1} n_{i j-1}-\sum_{i=1 }^{k} n_{ij}\geq 0
\quad\quad\quad 1\leq k < j\leq N  \label{nijrang}\ee
and
\be \sum_{j=i}^{k} n_{i-1 \, j-1}-\sum_{j=i}^{k} n_{ij} \geq
0 \quad\quad\quad 2\leq i \leq k \leq N \quad {\rm and} \quad i\leq N-1.
\label{nijlr} \ee
The weight $\mu$ of the second tableau and the weight $\nu$ of the resulting
LR tableau are easily recovered from these data.

The combined equations (\ref{nijrang}) and (\ref{nijlr}) constitute a set of linear and
homogeneous inequalities.  We call this the LR (or tensor-product) basis.
As described in \cite{Stan}, 
the Hilbert basis theorem guarantees that every solution can be
expanded in terms of the finite set of elementary solutions of these inequalities. This is a
key concept for the following (see \cite{BCM} for an extensive discussion of these methods). 
A sum of two solutions translates into the product of the corresponding couplings,
more precisely, to the {\it stretched product}  (denoted by $\cdot$) of the
corresponding two LR tableaux.  This is defined as follows.  Denote the empty
boxes of a LR tableau by a 0, so that
$ n_{0j} 
=\sum_{i=j}^{N-1} \lambda_i$, 
A tableau is thus  completely characterized by the data $\{n_{ij}\}$ where
now
$i\geq 0$. 
Then, the tableau obtained by the stretched product of the
tableaux  $\{n_{ij}\}$ and  $\{ n'_{ij}\}$ is simply described by the numbers 
$\{n_{ij}+n'_{ij}\}$, e.g.,
\be \matrix{\ST{\STrow{\bv\bv\b1}\STrow{\b1\b1\b2}\STrow{\b2\b3}\STrow{\b4}
}\cr} \cdot \matrix{\ST{\STrow{\bv\bv\b1}\STrow{\bv\b1\b2}\STrow{\bv\b2}
}\cr}  =  
\matrix{\ST{\STrow{\bv\bv\bv\bv\b1\b1}\STrow{\bv\b1\b1\b1\b2\b2}
\STrow{\bv\b2\b2\b3}\STrow{\b4} 
 }\cr}\quad  \ee

Let us now turn to the $su(2)$ case. 
The complete set of inequalities for $su(2)$ variables $\{\lambda_1, n_{11},
n_{12}\}$ is simply 
\be \lambda_1 \geq n_{12} \quad\quad
 n_{11}\geq 0 \quad\quad n_{12}\geq 0\label{inedeux} \ee
The first one expresses the fact that two boxes marked by a 1 cannot be in the
same column while the other two are obvious. The other weights are fixed by the
relation $
\mu_1 =n_{11}+n_{12}$ and $
\nu_1=\lambda_1+n_{11}-n_{12}.$
Any solution of these inequalities describes a coupling.
By inspection, the elementary solutions of this set of inequalities are
\be (\lambda_1, n_{11}, n_{12}) = (1,0,1), \quad (1,0,0), \quad (0,1,0)\ee
(For more complicated cases, we point out that powerful methods to find the
elementary solutions are described in \cite{BCM}.) These correspond to the
following LR tableaux, denoted respectively
$E_1, E_2, E_3$:
\be E_1: \quad \ST{\STrow{\bv}\STrow{\b1}}\, ,  
\quad\quad E_2: \quad \ST{\STrow{\bv}}\, ,
\quad\quad E_3: \quad \ST{\STrow{\b1}}  \ee
It is also 
manifest
that there are no linear relations between these couplings. Any stretched product
of these elementary tableaux is an allowed $su(2)$ coupling. Because there are no
relations between the elementary couplings, this decomposition is unique. We thus
see that the description of the elementary couplings captures, in a rather
economical way, the whole set of solutions of (\ref{inedeux}), that is, the whole set
of
$su(2)$ couplings.

Consider now the affine extension of these $su(2)$ results. The elementary
couplings have a natural affine extension, denoted by a hat, and their threshold
level is easily computed from the Kac-Walton formula.  The result is:
$k_0(\E_i)=1$ for
$i=1,2,3$. We observe that these values of  $k_0$ are the same as the number of
columns. Since the product of fusion elementary couplings is also a fusion and
because this decomposition is unique, we can read off the threshold level of any
coupling, hence of any LR tableau, simply from the number of its columns:
\be k_0 = \# {\rm columns} = \lambda_1+n_{11}\ee
And since $k$ is necessarily greater that $k_0$, we have obtained the extra
inequality:
\be k\geq \lambda_1+n_{11}\label{fudeux} \ee
This together with  (\ref{inedeux}) yield a set of inequalities describing completely
the fusion rules.  This is what we call a {\it fusion basis}, here the fusion
basis of
$\su(2)$. As in the finite case, the fusion couplings can be described in terms
of elementary fusions. These correspond to the elementary solutions of the four
inequalities, which are easily found to be 
\be (k,\lambda_1, n_{11}, n_{12}) = (1,0,0,0),\quad (1,1,0,1), \quad (1,1,0,0),
\quad (1,0,1,0)\ee
They correspond respectively to the couplings
\be  \eqalignD{ &\E_0 :[1,0]\times [1,0]\supset[1,0]\quad\quad \quad \E_2: [0,1]\times [1,0] \supset [0,1], \cr 
& \E_1 : [0,1]\times [0,1] \supset [1,0] \quad\quad\quad  \E_3 : [1,0]\times [0,1] \supset [0,1]. \cr} \ee 

Any fusion has an unique decomposition in terms of these
elementary couplings. For instance
\be [3,2]\times[1,4]\supset[1,4] \quad \lra \quad
\ST{\STrow{\bv\bv\b1\b1\b1}\STrow{\b1}}\quad \lra \quad \E_1\E_2\E_3^3:\quad
k_0=5\ee

\section{Constructing the fusion basis}

For algebras other than $\su(2)$, the threshold level is not simply the number of
columns.  So the question is:  how can we derive the fusion basis? The strategy,
developed in \cite{BCMa} is the following: 

{\bf 1}- write the LR inequalities; 

{\bf 2}- from these, find the tensor-product elementary couplings; 

{\bf 3}- from these, find fusion elementary couplings;

{\bf 4}- from these, reconstruct the fusion basis.

To go from step 2 to step 3, we need some tools; we describe below a method
based on the outer automorphism group.  Similarly to go from 3 to 4, we need
a further ingredient: this is the Farkas' Lemma.  We discuss these techniques in
turn.

Let us start from the set of tensor-product elementary couplings $\{E_i, i\in I\}$
for some set $I$ fixed by the particular $su(N)$ algebra under study. For each
$E_i$, we  calculate the threshold level $k_0(E_i)$ and this
datum specifies the affine extension of
$E_i$,  denoted $\E_i$.  We have then a partial set of fusion elementary couplings
with the set
$\{\E_i,i\in I\}$. Our conjecture is that the missing fusion elementary couplings
can all be generated by the action of the outer-automorphism group. For
$\su(N)$, this group is simply $\{a^n ~|~ n=0,\cdots, N-1\}$, with
\be a[\lambda_0,\lambda_1,\cdots,\lambda_{N-1}]=
[\lambda_{N-1},\lambda_0,\cdots,\lambda_{N-2}]\ee The conjecture is based on the
invariance relation 
\be  {\Nc_{a^n\lah, a^m\muh}^{(k)}}^{~a^{n+m}\nuh}=
{\Nc_{\lah\muh}^{(k)}}^{~\nuh}\ee 
and it amounts to supposing that the full set is
contained in

\be \{\E_i (a^n \lah, a^m \muh, a^{n+m} \nuh)~|~i \in I, ~0 \leq m,n <N\} \ee


\noindent The conjectured completeness requires the consideration of all possible pairs
$(n,m)$.\footnote[1]{Note that we do not suppose that the action of ${\cal A }$ on an
elementary coupling will necessarily produce another elementary coupling.  Indeed,
the resulting coupling could be a product of elementary couplings.  What is
conjectured here is that all fusion elementary couplings can be generated in this
way.}

Let us illustrate this with the $\su(2)$ case. Start with the elementary
coupling $E_1:\, (1)\otimes (1)\supset (0)$, which, as already indicated, 
arises at level 1: $k_0(E_1)=1$. The corresponding fusion is thus $[0,1]\times
[0,1] \supset [1,0]$,  denoted as $\E_1$.
We now consider  all possible actions of the outer-automorphims group on it. 
Since this group is of order 2, there are 4 possible choices for the pair
$(n,m)$:
\be (a^n,a^m)\in \{(a,a),(1,1), (1,a), (a,1)\}\ee
with $a[\lambda_0, \lambda_1] = [\lambda_1, \lambda_0]$.
This generates the set of four elementary couplings found previously, in the order $\E_0,\E_1,\E_2,\E_3$. Thus, from one tensor-product elementary
coupling, all four fusion elementary couplings are deduced. 

We now turn to Farkas' lemma.  For its presentation, it is convenient to use
an exponential description of the couplings, that is, 
\be (k,\lambda_i,n_{ij})\quad \rw \quad d^k L_i^{\lambda_i}N_{ij}^{n_{ij}}\ee
 $d,\,L_i,\,N_{ij}$ being dummy variables.  For instance $\E_1$ is
represented by
$dL_1N_{12}$. If we collectively describe a coupling by the complete set of
variables $\{x_i\}$, we have
\be \{x_i\} \quad \rw \quad \{X_i^{x_i}\}\ee
A particular coupling is thus described by a given product $\prod_i
X_i^{x_i}$ with fixed $x_i$.  Its decomposition in terms of elementary
couplings  takes the form $\prod_i \E_i^{\,a_i}$.  Now, since $\E_i$ can be
decomposed in terms of the $X_j$ as \be \E_i = \prod_j
X_j^{\e_{ij}}\label{epde} \ee it means that reading off a particular coupling
means that we are interested in a specific choice of the set of positive integers
$\{x_i\}$ fixed by
\be \sum_i a_i\e_{ij}  = x_j\ee
in terms of some nonnegative integers $a_i$. 
We are thus looking for the existence
conditions for such a coupling, i.e., the underlying set of linear and
homogeneous inequalities. This is exactly what the Farkas' Lemma \cite{Schri, Bek} provides: given the
knowledge of the elementary couplings, it allows us 
to recover the underlying set of
inequalities. For tensor
products, this is of no interest since 
we know the corresponding set of inequalities
and our elementary couplings have been extracted from them.  But the situation is
quite different in the fusion case, where the {\it fusion basis} is unknown.

For our application we need the following modification of the lemma, proved in
\cite{BCMa}:

\n {\it Lemma}: 
Let $A$ be an $r\times m$ integer matrix and let
$\epsilon_j$, $j=1\dots n$ be a set of fundamental
solutions to 
\be Ax\geq 0,\quad x\in {\bf N}^m. \label{condB} \ee 
${\bf N}=\{0,1,2,\dots\}$. Let
$V$ be the $m\times n$ matrix with entries $V_{i,j}=(\epsilon_j)_i$
$i=1\dots m$, $j=1\dots n$, i.e., the columns
of $V$ are a set of fundamental solutions to (\ref{condB}).
Let $e_w$, $w=1\dots k$ be a fundamental system
of solutions of 
$ u^\top V\geq 0$
(not necessarily positive) $ u\in {\bf Z}^m$ 
and let $E$ be the $k\times m$ matrix
with entries $E_{w,i}=(e_w)_i$, i.e.~ the rows
of $E$ are the fundamental solutions $e_w$, $w=1\dots k$.
Then the solution set of the system
\be Ex\geq 0,\quad x\in {\bf N}^m \label{condC} \ee
is the same as the solution set of (\ref{condB}).

To link  the lemma to the situation presented  above, we note that the entries
$V_{ij}$ of the matrix
$V$ are given here by the numbers $\e_{ji}$ appearing in (\ref{epde}).
  Our analogue
of the relation
$V\, a=x$ describes a generic coupling and our goal is to find the defining
system of inequalities underlying the existence of this coupling.

Take a simple example, the $\su(2)$ case. The
elementary couplings and the corresponding vectors $\e_i$ are
\be \eqalignD{ &\E_0 : d\quad\quad & \e_0= (1,0,0,0)\cr
&\E_1: dL_1N_{12}\quad\quad  & \e_1= (1,1,0,1)\cr
&\E_2: dL_1\quad\quad  & \e_2= (1,1,0,0)\cr
&\E_3: dN_{11}  & \e_3= (1,0,1,0)\cr}\label{eleuu} \ee
 From the vectors $\e_{i}$ with components $\e_{ij}$, we form
the matrix $V$ with entries $V_{ij}=\e_{ji}$:
\be V= \pmatrix {1&1&1&1\cr 0&1&1&0\cr 0&0&0&1\cr 0&1&0&0\cr}\ee
With $a$ and $x$ denoting the column matrices of entries $a_i$ and $x_i$
respectively, we have the matrix equation
\be V\, a=x\ee
This equation describes a general fusion coupling.  We now want to unravel the
underlying system of inequalities.  For this, we consider the fundamental solutions
of 
\be u^\top\, V\geq 0\ee where $u$ is the vector of entries $u_i$.
These inequalities read
\be \eqalignD{ & u_0\geq 0\cr
& u_0+u_1+u_3\geq 0\cr
& u_0+u_1\geq 0\cr
& u_0+u_2\geq 0\cr} \ee
In this simple case, the elementary couplings can be found by inspection and these
are:
\be e_0=(1,-1,-1,0),\quad\quad e_1=(0,0,0,1),\quad\quad 
e_2=(0,1,0,-1),\quad\quad 
 e_3=(0,0,1,0)\ee
  Finally, we consider the conditions
$e_i\, x\geq 0$, with
$(x_0,x_1,x_2,x_3)= (k,
\lambda_1, n_{11}, n_{12})$.  They read, in order,
\be k\geq \lambda_1+n_{11},\quad\quad\quad
n_{12}\geq 0,\quad\quad\quad
\lambda_1\geq n_{12},\quad\quad\quad
n_{11}\geq 0 \ee
The last three conditions define the LR basis.  The first one is the additional
fusion constraint.  Together they form the $\su(2)$ fusion basis.

\section{Conclusion}

This method has been extended to other algebras in \cite{BCMa} and worked out in
detail for
$\su(3,4)$ and $\sp(4)$.  This leads to explicit expressions for the threshold
levels, hence for fusion coefficients.   For algebras other that
$\su(N)$, we replace the LR basis by the Berenstein and Zelevinsky basis \cite{BZin}.

Clearly, the main open problem is to find 
a fundamental and
Lie algebraic way of deriving the basis, analogous in spirit to the
Berenstein-Zelevinsky conjectures for generic Lie algebras in \cite{BZin}.

Finally, we stress that the reformulation of the problem of computing fusion rules
in terms of a fusion basis solves, in principle, the quest for
a combinatorial method since it  reduces a fusion computation to solving
inequalities. But this is probably  not the optimal solution to the quest
for an efficient combinatorial description.

\section{Acknowledgments}

This work was supported by NSERC (Canada) et FCAR (Qu\'ebec).

\end{document}